\documentclass[twocolumn,showpacs,preprintnumbers,amsmath,amssymb]{revtex4}

\usepackage{graphicx}
\usepackage{dcolumn}
\usepackage{bm}

\newcommand{\cuncs}{$\kappa$-(ET)$_2$Cu(NCS)$_2$}

\newcommand{\tc}{$T_c$}
\newcommand{\gr}{$\kappa$}
\newcommand{\ces}{$C_{\rm es}$}
\newcommand{\cph}{$C_{\rm ph}$}
\newcommand{\cn}{$C_{\rm n}$}

\begin{document}
\title{High-resolution ac-calorimetry studies of the quasi-2D
organic superconductor $\kappa$-(BEDT-TTF)$_2$Cu(NCS)$_2$}
\author{J.\ M\"uller}
 \email{mueller@cpfs.mpg.de}
\affiliation{Max-Planck-Institut f\"ur Chemische Physik fester Stoffe, N\"othnitzer-Str.
40, D-01187 Dresden, Germany}
\author{M.\ Lang}
\affiliation{Physikalisches Institut der J.\ W.\ Goethe-Universit\"at Frankfurt am Main,
FOR 412, D-60054 Frankfurt am Main, Germany}
\author{R.\ Helfrich, F.\ Steglich}
\affiliation{Max-Planck-Institut f\"ur Chemische Physik fester Stoffe, Dresden, Germany}
\author{T.\ Sasaki}
\affiliation{Institute for Materials Research, Tohoku University, Sendai, Japan}

\date{\today}

\begin{abstract}
We report high-resolution specific heat measurements on the quasi-2D organic superconductor
$\kappa$-(BEDT-TTF)$_2$Cu(NCS)$_2$, performed by using a most sensitive ac-modulation
technique. The main observations are (i) a discontinuity at \tc\ much in excess of what is
expected for a weak-coupling BCS superconductor and (ii) a quasiparticle contribution to
the specific heat with an exponentially weak temperature dependence at $T \ll T_c$. The
latter finding is incompatible with an order parameter that vanishes at certain parts of
the Fermi surface. The data for $T \leq T_c$ can be well described by a strong-coupling
extension of the BCS theory - the $\alpha$-model - similar to what has been recently found
for the $\kappa$-(BEDT-TTF)$_2$Cu[N(CN)$_2$]Br salt [Elsinger {\em et al.}, Phys. Rev.
Lett. {\bf 84}, 6098 (2000)].
\end{abstract}


\maketitle



The quasi-twodimensional organic charge-transfer salts \gr-(BEDT-TTF)$_2$X based on the
electron-donor molecule bis(ethylenedithio)-tetrathiafulvalene, commonly abbreviated to
BEDT-TTF or simply ET, are of particular interest because of their rich phase diagram
including superconductivity at relatively high temperatures next to an antiferromagnetic
insulating state \cite{Ishiguro98,Kanoda97}. The pressure-induced transition from an
antiferromagnetic insulator to a superconductor for X = Cu[N(CN)$_2$]Cl is one of the
intriguing observations that places these materials in line with other exotic
superconductors where pairing mechanisms which are different from the conventional
electron-phonon interaction might be of relevance. In fact, the proximity of magnetic order
and superconductivity along with the presence of spin fluctuations above \tc\ as inferred
from $^{13}$C-NMR measurements \cite{Mayaffre94,Kawamoto95,de Soto95}, have been taken as
strong indications for a spin-fluctuation-mediated superconductivity
\cite{Kino98,Kondo98,Kuroki99,Schmalian98,Louati00} similar to that which has been proposed
first for the high-$T_c$ cuprates \cite{Scalapino87}. Since the identification of the
relevant pairing interaction is a very difficult problem, many experiments have focussed on
the determination of the structure of the superconducting order parameter - the gap
amplitude $\Delta$(\boldmath $k$\unboldmath) - which is intimately related to the pairing
mechanism. However, despite intensive experimental efforts devoted to this issue, no
consensus has been achieved yet \cite{Lang96}. Arguments in favour of an unconventional
order-parameter with d-wave symmetry for the above \gr-(ET)$_2$X salts have been derived
from temperature-dependent investigations, notably NMR measurements \cite{de
Soto95,Mayaffre95,Kanoda96}, thermal conductivity \cite{Belin98} and one specific heat
study \cite{Nakazawa97}. More recent attempts such as STM \cite{Arai01}, mm-wave
transmission \cite{Schrama99} or thermal conductivity \cite{Izawa01} have focussed on
orientational-dependent investigations aiming at a direct determination of the gap
anisotropy. Although a modulation of the gap structure has been seen in all of these
studies, they come to quite different conclusions on the direction of the gap zeroes, see
e.g.\ \cite{Izawa01}.
\newline
Conversely, there are numerous experimental studies that are consistent with a conventional
BCS-type of superconductivity. Among them are measurements of the magnetic penetration
depth \cite{Lang92} and surface impedance \cite{Klein91} as well as the observation of a
BCS-like mass isotope effect \cite{Kini96} and a pronounced superconductivity-induced
phonon renormalization \cite{Pintschovius97}. The latter two experiments clearly
demonstrate the relevance of intermolecular phonons for the pairing interaction.
\newline
A very powerful method to probe certain aspects of the gap structure - in particular the
question whether gap zeroes exist or not - is provided by specific heat measurements. In
case this integral thermodynamic technique were to find a low temperature electronic
quasiparticle contribution, $C_{\rm es}$, that varies exponentially weakly with the
temperature, the existence of gap zeroes on the Fermi surface could be definitely ruled
out. On the other hand, the observation of a non-exponential temperature dependence does
not represent a clear proof of gap zeroes as it may also originate in extraneous
contributions such as impurity phases, normal-conducting regions or pair breaking. In fact,
an exponential low-T specific heat behaviour implying a finite energy gap has been found in
recent high-resolution specific heat measurements on the X = Cu[N(CN)$_2$]Br salt
\cite{Elsinger00}. Moreover it has been shown in this study that the $T^2$ dependence in
\ces\ reported for the same compound by Nakazawa et al.\ \cite{Nakazawa97} most likely
originates in their incorrect determination of the phonon background.
\newline
Here we present high-resolution specific heat measurements on the related
\cuncs\ salt with the intention to clarify the existence or absence of gap
nodes for this compound.

The single crystals used were synthesised by the standard electrocrystallization technique
as described elsewhere \cite{Sasaki90}. For the present investigations, two high-quality
single crystals with regular plate-like shapes and masses of 0.63\,mg (\#\,1) and 0.72\,mg
(\#\,2) have been selected. The specific heat has been measured utilising a high-resolution
ac-modulation technique \cite{Sullivan68}. The setup has been designed for investigating
very small plate-like single crystals such as the present compounds. The sample holder
consisting of a resistive thermometer (Cernox CX-1080-BG) and heater is attached to a
$^4$He-bath cryostat equipped with an 8\,T superconducting solenoid. The calorimeter has
been checked by measuring high-purity samples of Cu and Ag with typical masses of about
$3\,\sim\,4$\,mg which have absolute heat capacities comparable to those of the small
crystals studied here. In the temperature range $2\,{\rm K} \leq T \leq 30$\,K, the maximum
deviations from the literature results amount to $\pm 2$\,\%.

\begin{figure}[tbp]
\includegraphics[width=8.5cm]{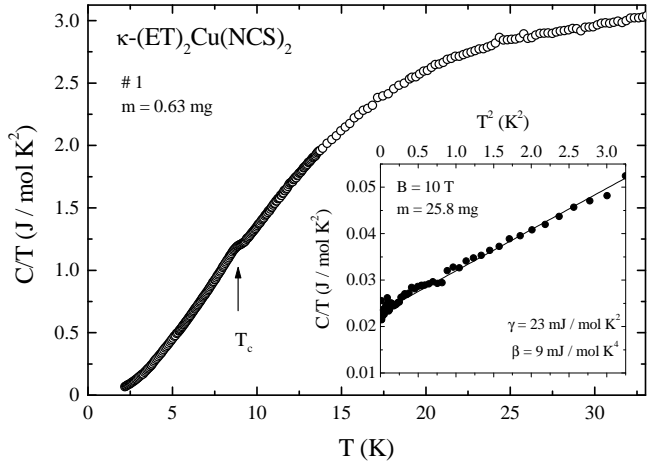}
\caption{Specific heat data as $C/T$ {\em vs} $T$ of \cuncs, crystal \#\,1. The inset shows
data as $C/T$ {\em vs} $T^2$ on a compilation of several single crystals with a total mass
of $25.8$\,mg taken at low temperatures in a magnetic field $B = 10$\,T. The solid line is
a linear fit of the form $C/T = \gamma + \beta T^2$ to the data below $T=2$\,K.}
\label{fig1}
\end{figure}

\begin{figure}[tbp]
\includegraphics[width=6cm]{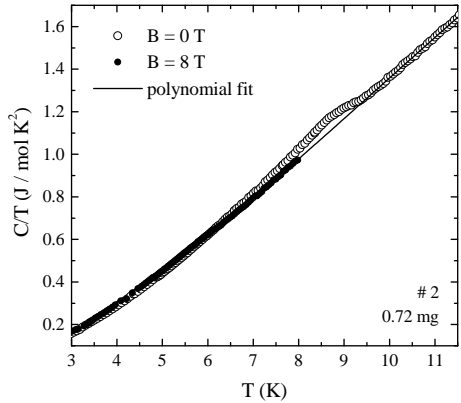}
\caption{Specific heat data as $C/T$ {\em vs} $T$ of \cuncs, crystal \#\,2, in the vicinity
of the superconducting transition. Data have been taken in zero field and in an
overcritical field $B=8$\,T. The line represents a polynomial fit to the data $C(T \leq
8\,{\rm K},B=8\,{\rm T})$ and $C(T \geq 10\,{\rm K},B=0)$, see text.} \label{feld}
\end{figure}

\begin{figure}[tbp]
\includegraphics[width=8.5cm]{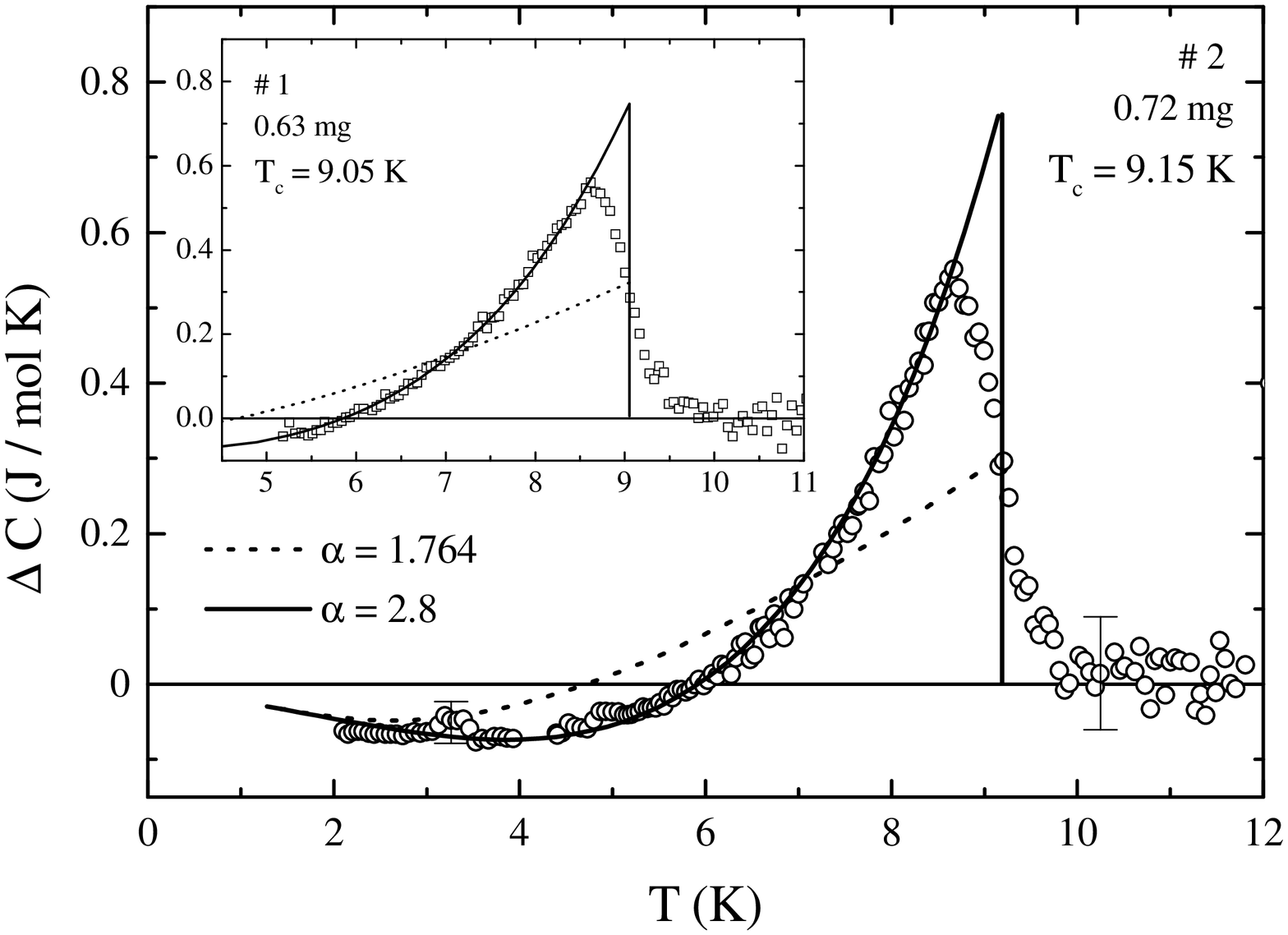}
\caption{Specific heat difference $\Delta C=C(0\,{\rm T})- C(8\,{\rm T})=C(0\,{\rm T})-
C_{\rm n}$ of crystals \#\,2 and \#\,1 (inset). The dotted and solid thick lines represent
the BCS curves for weak and strong coupling, respectively.}\label{fig2}
\end{figure}

\begin{figure}[tpb]
\includegraphics[width=6cm]{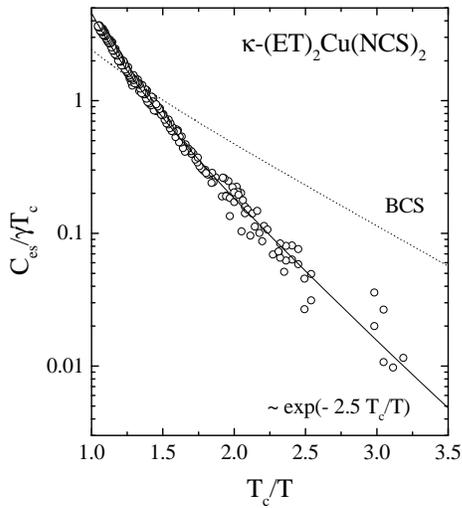}
\caption{Semi-logarithmic plot of the electronic contribution to the specific heat in the
superconducting state as $C_{\rm es}/\gamma T_c$ {\em vs} $T_c/T$. The dotted and solid
lines represent the weak- and strong-coupling BCS behaviour, respectively, see
text.}\label{bcs}
\end{figure}

Fig.\ \ref{fig1} shows the specific heat as $C/T$ {\em vs} $T$ of crystal
\#\,1 over the entire temperature range investigated \cite{comment0}. The
phase-transition anomaly at \tc\ is clearly visible although it amounts to
only about five percent of the total specific heat at this temperature. To
determine the quantity of interest - the quasiparticle specific heat in the
superconducting state \ces\ - one has to get rid of the large phonon
background \cph. It is known for these molecular systems that, due to
low-lying optical phonon modes, \cph\ starts to deviate from a Debye-like
behaviour already at low temperatures \cite{Wosnitza94,Wanka98}. To avoid
uncertainties related to \cph, we proceed in analyzing the difference
$\Delta C(T) = C(T,B=0) - C(T,B=8\,{\rm T})$. For the field orientation used
in our experiment, $B
\parallel a^\ast$, where the upper critical field is $B_{c_2} \approx 6$\,T
\cite{Lang94}, the data taken at 8\,T represent the normal-state specific heat \cn.
Provided that \cn\ is field independent, which has been found in previous measurements for
the present compound \cite{Andraka89} and also for X = Cu[N(CN)$_2$]Br \cite{Elsinger00},
the quantity $\Delta C$ has the advantage that the unknown \cph\ and all other extraneous
contributions (such as a small amount of Apiezon grease used to improve the thermal contact
to the thermometer and heater) cancel each other out. The experimental fact that \cn\ is
independent of the magnetic field within the experimental resolution
\cite{Andraka89,Elsinger00} is consistent with $C_{\rm n} = C_{\rm ph} + \gamma T$
\cite{comment1}, where $\gamma$ is the Sommerfeld coefficient.
\newline
Fig.\ \ref{feld} shows the specific heat data for $B = 0$ and $B = 8$\,T applied along the
$a^\ast$-axis, i.e.\ perpendicular to the conducting planes. For technical reasons,
measurements in fields of $B = 8$\,T were limited to $T \leq 8$\,K. For $C(T,B=8\,{\rm T})
= C_{\rm n}(T)$ at $T \geq 8$\,K, an interpolation based on a polynomial fit between the
$C(T < 8\,{\rm K},B = 8\,{\rm T})$ and $C(T > 10\,{\rm K},B=0)$ data was used, cf.\ the
solid line in Fig.\ \ref{feld}. The so-derived $\Delta C = C(0\,{\rm T}) - C(8\,{\rm T}) =
C_{\rm es} - \gamma T$ of crystal \#\,2 is shown in Fig.\ \ref{fig2} together with $\Delta
C$ expected from the BCS weak-coupling theory \cite{Muehlschlegel59}. The theoretical curve
is based on a Sommerfeld coefficient $\gamma = (23 \pm 1)$\,mJ/mol\,K$^2$ as determined by
low-temperature $C(T)$ measurements employing a thermal-relaxation technique on a
compilation of three single crystals with total mass of 25.8\,mg, see inset of Fig.\
\ref{fig1} \cite{comment2}. This $\gamma$ value is consistent with $\gamma = (25 \pm
3)$\,mJ/mol\,K$^2$ derived from an earlier specific heat study on the same compound
\cite{Andraka89} by analysing the data in the range $1.3\,{\rm K} < T < 3$\,K which is
slightly above that used here \cite{comment3}.
\newline
Fig.\ \ref{fig2} demonstrates that, similarly to what has been recently observed for the X
= Cu[N(CN)$_2$]Br salt \cite{Elsinger00}, $\Delta C(T)$ deviates markedly from the
weak-coupling BCS-behaviour in both the jump height at \tc\ as well as the overall
temperature dependence. However, as for X = Cu[N(CN)$_2$]Br, a much better description of
the data is obtained by using the semi-empirical extension of the BCS formalism to
strong-coupling superconductors - the so-called $\alpha$-model \cite{Padamsee73}. It
contains a single free parameter $\alpha \equiv \Delta(0)/k_B T_c$ ($k_B$ being the
Boltzmann constant) which scales the BCS energy gap $\Delta (T) = (\alpha / \alpha_{\rm
BCS}) \cdot \Delta_{\rm BCS}(T)$ with $\alpha_{\rm BCS}=1.764$. As Fig.\ \ref{fig2} clearly
demonstrates, the strong-coupling BCS model with $\alpha = 2.8 \pm 0.1$ provides an
excellent description of the data over the entire temperature range investigated. The error
margins in $\alpha$ account for the uncertainties in both $\gamma$ as well as \tc. In the
inset of Fig.\ \ref{fig2} we show $\Delta C$ data of crystal \#\,1 together with the
theoretical results of the $\alpha$ model. Although \tc\ of crystal \#\,1 is slightly
reduced compared to that of crystal \#\,2, we find again an excellent agreement with the
strong-coupling results employing the same parameter $\alpha = 2.8$.
\newline
The fact that both data sets in Fig.\ \ref{fig2} are well described within the
strong-coupling BCS model implies (i) that \ces\ reveals an exponentially weak temperature
dependence at low temperatures and (ii) the thermodynamic consistency of the data, i.e.\
entropy conservation. The exponential variation of \ces\ becomes clearer in Fig.\
\ref{bcs}, where $C_{\rm es}/\gamma T_c$ is shown in a semi-logarithmic plot as a function
of $T_c/T$. Here, $C_{\rm es}=\Delta C + \gamma T$ with $\gamma = (23 \pm
1)$\,mJ/mol\,K$^2$ has been used. The solid line represents the same strong-coupling curve
shown in Fig.\ \ref{fig2}. The figure also includes the weak-coupling BCS result (dashed
line) which predicts $C_{\rm es}/\gamma T_c \propto \exp(-a_{\Delta} T_c / T)$ with
$a_\Delta = 1.44$ for $2.5\,\leq T_c/T \leq 6$ \cite{Gladstone69}. The data of \cuncs\
presented here follow $C_{\rm es}/\gamma T_c \propto \exp(-2.5 T_c / T)$ down to the lowest
accessible temperature. The enhanced prefactor in the exponent of $a_\Delta = 2.5$ as
compared to 1.44 for the weak-coupling BCS model reflects the strong-coupling character of
the present superconductor.
\newline
Fig.\ \ref{bcs} demonstrates that our results are fully consistent with an exponentially
vanishing \ces\ at low temperatures and, thus, an energy gap without zeroes at the Fermi
surface. The same conclusions have been drawn from a similar analysis of $C(T,B)$ data on
\gr-(ET)$_2$Cu[N(CN)$_2$]Br \cite{Elsinger00}. As the specific heat is an integral
technique that picks up all excitations at the Fermi surface, the above findings are
incompatible with the existence of gap zeroes as claimed by other experiments, notably NMR
\cite{de Soto95,Mayaffre95,Kanoda96}. Since the latter results have been obtained in a
finite magnetic field applied parallel to the conducting planes, the discrepancies are
possibly related to an as yet not understood influence of the magnetic field.
\newline
In conclusion, high-resolution specific heat measurements on small high-quality single
crystals of \gr-(BEDT-TTF)$_2$Cu(NCS)$_2$ have been performed in both the normal and
superconducting state. The data analysis employed, which minimises uncertainties associated
with the unknown, large phonon background reveals an electronic quasiparticle contribution
to the specific heat that varies exponentially weakly with temperature at $T \ll T_c$. This
behaviour is fully consistent with a finite energy gap all over the Fermi surface and rules
out the existence of gap nodes. Moreover, we find an excellent agreement with the
predictions of the strong-coupling variant of the BCS model employing an $\alpha$ parameter
2.8 that slightly exceeds the one found recently for the related
\gr-(ET)$_2$Cu[N(CN)$_2$]Br salt \cite{Elsinger00}.

\vspace{4cm}

\begin{thebibliography}{37}
\bibitem{comment0} Note that due to the small absolute heat capacities of
the tiny crystals studied reliable data could be obtained only for $T \geq
2$\,K.
\bibitem{comment1} A field independent $C_{\rm n}$ implies that possible other contributions such
as, e.g.\ due to spin fluctuations are either very small (or even absent) or
field independent up to the field level investigated.
\bibitem{comment3}Due to a significant non-superconducting volume fraction of about 10\,\%
found in these low-temperature studies on the composite samples with $m = 25.8$\,mg which
most likely originates in the irregular surface of one of the crystals and the pressure
exerted by frozen-in Apiezon grease, a reliable determination of the low-temperature \ces\
was impeded. This, however, does not affect the determination of $\gamma$ within the error
margins specified.
\end{thebibliography}
\end{document}